\documentclass[prl,twocolumn,showpacs,superscriptaddress]{revtex4-1} 
\usepackage{amsmath,amssymb,graphicx}
\begin{document}

\title{Optical pump-rejection filter based on silicon sub-wavelength engineered photonic structures}

\author{Diego P\'{e}rez-Galacho}\thanks{These two authors contributed equally.}
\affiliation{Centre for Nanoscience and Nanotechnology, CNRS, Univ. Paris-Sud, Universit\'e Paris-Saclay, C2N – Orsay, 91405 Orsay cedex, France}
\author{Carlos Alonso-Ramos}\thanks{These two authors contributed equally.}
\affiliation{Centre for Nanoscience and Nanotechnology, CNRS, Univ. Paris-Sud, Universit\'e Paris-Saclay, C2N – Orsay, 91405 Orsay cedex, France}
\author{Florent Mazeas}
\affiliation{Universit\'{e} C\^{o}te d'Azur, CNRS, Institut de Physique de Nice, Parc Valrose, 06108 Nice Cedex 2, France}
\author{Xavier Le Roux}
\affiliation{Centre for Nanoscience and Nanotechnology, CNRS, Univ. Paris-Sud, Universit\'e Paris-Saclay, C2N – Orsay, 91405 Orsay cedex, France}
\author{Dorian Oser}
\affiliation{Centre for Nanoscience and Nanotechnology, CNRS, Univ. Paris-Sud, Universit\'e Paris-Saclay, C2N – Orsay, 91405 Orsay cedex, France}
\author{Weiwei Zhang}
\affiliation{Centre for Nanoscience and Nanotechnology, CNRS, Univ. Paris-Sud, Universit\'e Paris-Saclay, C2N – Orsay, 91405 Orsay cedex, France}
\author{Delphine Marris-Morini}
\affiliation{Centre for Nanoscience and Nanotechnology, CNRS, Univ. Paris-Sud, Universit\'e Paris-Saclay, C2N – Orsay, 91405 Orsay cedex, France}
\author{Laurent Labont\'{e}}
\affiliation{Universit\'{e} C\^{o}te d'Azur, CNRS, Institut de Physique de Nice, Parc Valrose, 06108 Nice Cedex 2, France}
\author{S\'{e}bastien Tanzilli}
\affiliation{Universit\'{e} C\^{o}te d'Azur, CNRS, Institut de Physique de Nice, Parc Valrose, 06108 Nice Cedex 2, France}
\author{\'{E}ric Cassan}
\affiliation{Centre for Nanoscience and Nanotechnology, CNRS, Univ. Paris-Sud, Universit\'e Paris-Saclay, C2N – Orsay, 91405 Orsay cedex, France}
\author{Laurent Vivien}
\affiliation{Centre for Nanoscience and Nanotechnology, CNRS, Univ. Paris-Sud, Universit\'e Paris-Saclay, C2N – Orsay, 91405 Orsay cedex, France}

\begin{abstract}
The high index contrast of the silicon-on-insulator (SOI) platform allows the realization of ultra-compact photonic circuits. However, this high contrast hinders the implementation of narrow-band Bragg filters. These typically require corrugations widths of a few nanometers or double-etch geometries, hampering device fabrication. Here we report, for the first time, on the realization of SOI Bragg filters based on sub-wavelength index engineering in a differential corrugation width configuration. The proposed double periodicity structure allows narrow-band rejection with a single etch step and relaxed width constraints. Based on this concept, we experimentally demonstrate a single-etch, $\mathbf{220\,nm}$ thick, Si Bragg filter featuring a corrugation width of $\mathbf{150\,nm}$, a rejection bandwidth of $\mathbf{1.1\,nm}$ and an extinction ratio exceeding $\mathbf{40\,dB}$. This represents a ten-fold width increase compared to conventional single-periodicity, single-etch counterparts with similar bandwidths.\\
\textit{OCIS: (130.3120) Integrated optics devices; (230.7390) Waveguides, planar; (230.1480) Bragg reflector; (040.6040) Silicon.}
\end{abstract}


\maketitle 

The silicon-on-insulator (SOI) platform with sub-micrometric thick Si layer has shown outstanding results in the miniaturization of photonic circuits \cite{SiPhot}. High-quality materials and mature fabrication processes, together with the potential to leverage already existing CMOS facilities, make it a promising candidate for the large volume production of performant photonic devices. In addition to datacom \cite{DataCom} or sensing applications \cite{SensingA,SensingB}, SOI shows a great potential for the generation and manipulation of photonic entanglement \cite{SiQuantumA,SiQuantumB,SiQuantumC,SiQuantumD,SiQuantumE,SiQuantumF}. Such a technology would enable monolithic integration of quantum-processing circuits, opening new routes for envisioned quantum-based applications, including quantum key distribution  \cite{QuantumKD} and optical quantum computing \cite{QuantumComputing}. 

	\begin{figure}[htbp]
		\centering
		\centerline{\includegraphics[width=\columnwidth]{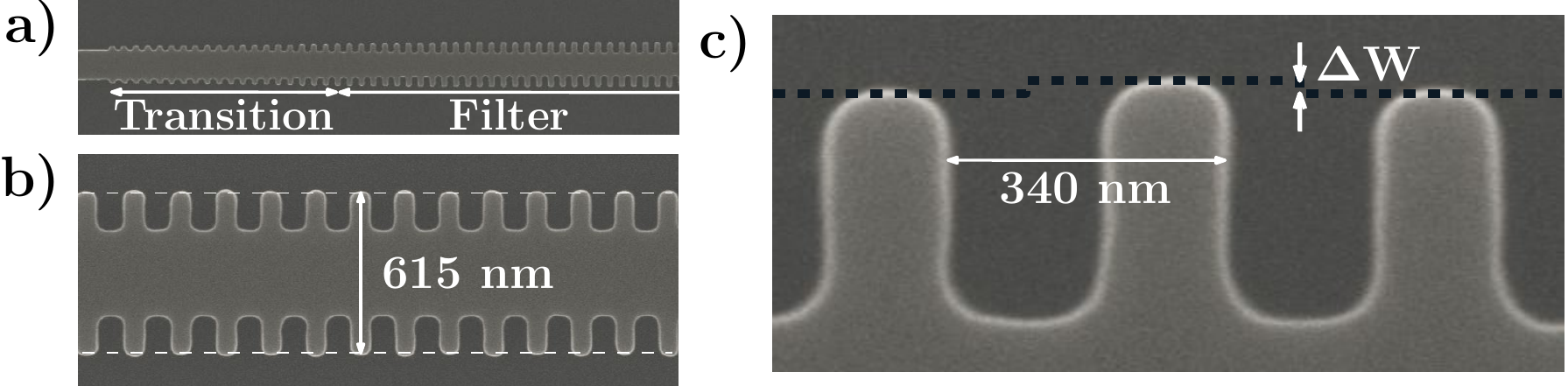}}
		\caption{Scanning electron microscope (SEM) images of fabricated filters: (a) adiabatic transition between strip waveguide and filter, (b) filter and (c) detail of double-periodicity structure.}
		\label{fig:FabrSEMs}
	\end{figure}
	
	Spontaneous four-wave mixing in Si micro-resonators has already demonstrated efficient on-chip generation of entangled photon pairs \cite{SiQuantumA,SiQuantumC,SiQuantumD,SiQuantumF}. Owing to the substantially higher pump intensity, compared to that of the photon-pair signal, on-chip pump-rejection filters are essential for integrated quantum circuits. Besides strong pump suppression, narrow rejection bandwidth is particularly important to allow short wavelength separation of the paired photons, thereby minimizing dispersion effects in the micro-ring that may reduce the efficiency of the nonlinear process. 
	
	Due to the high-index contrast in SOI, the implementation of narrow-band, as well as high-rejection SOI filters is a real challenge. In this work, we report the design and experimental characterization of novel sub-wavelength engineered Bragg filters that overcome this limitation, simultaneously showing narrow band operation and high rejection level. Remarkably narrow-band filters with sub-nanometer wide rejections have been previously reported, based on sophisticated architectures that combine micro-ring resonators and reflectors \cite{Yo_Melloni} or contra-directional couplers \cite{RR_ContrDirCoup}. Modulation of the waveguide cladding has also been used to realize narrow-band filters \cite{CladdMod_A,CladdMod_B}. Nevertheless, these solutions exhibit modest rejection levels that preclude their use as pump-rejection filters. Ultra high-rejection filters, based on cascaded Mach–Zehnder interferometers have recently been reported \cite{MZI_Filter}. However, they require active tuning and exhibit a comparatively large rejection bandwidth. On the other hand, high-rejection Si Bragg filters can be straightforwardly realized by judiciously modulating the waveguide width. Still, the very large index contrast offered by the SOI platform hinders the implementation of narrow bandwidths. Indeed, corrugation widths of only 10 nm can be required \cite{Small_Teeth}, which complicates device fabrication. Bragg filters relying on shallow-etched rib geometries have been reported with narrow rejection bandwidth and relaxed widths (exceeding 80 nm) \cite{Rib_a,Rib_b}. However, they require a two-step fabrication process that compromises the cost-effectiveness of this solution. Recently, very promising contra-directional cross-mode coupling in asymmetrically corrugated multi-mode Si waveguides has been used for narrow-band rejection with corrugation widths larger than $100\,\mathrm{nm}$ \cite{AsymBragg}. The transverse-magnetic (TM) polarized mode in the Si-wire can also be considered, as it is less confined than the transverse-electric (TE) mode, thus resulting in comparatively lower effective indices. Hence, TM Bragg filters enable bandwidth of $\sim 1 \,\mathrm{nm}$ with corrugation widths of $60 \,\mathrm{nm}$ \cite{Spiral_TM}. Alternatively, high-index-contrast constraints in the SOI platform can be overcome by waveguide index engineering based on sub-wavelength structuration \cite{FirstSwg}. Periodically patterning the waveguide with a pitch smaller than half of the propagating wavelength, makes it possible to obtain an arbitrary effective index between those of the Si and the cladding material \cite{SwgReview,SwgBraggSim,SwgSimul}. Bragg filters, relying on sub-wavelength index contrast engineering in fully segmented waveguides, have experimentally demonstrated narrow operation (3 dB bandwidth of only $0.5 \,\mathrm{nm}$) with a moderate rejection level of $12\,\mathrm{dB}$ \cite{SwgBraggExp}.

	Here, we propose a sub-wavelength engineered Bragg filter geometry that allows single etch step and relaxed minimum feature size constraints. In the proposed filter geometry, shown in Fig. \ref{fig:FabrSEMs}, we divide the Bragg period in two sub-wavelength periods with slightly different corrugation widths. Thus, the Bragg modulation strength is mainly determined by the difference between the widths of the two sub-wavelength corrugations, rather than by their absolute width. We exploit this new degree of freedom to implement high-rejection and narrow-band Bragg filters with relaxed requirements on minimum corrugation width. Based on this novel geometry, we implement Bragg filters in the SOI platform with a 220 nm thick Si layer, showing a bandwidth of $1.1\,\mathrm{nm}$ and a rejection level exceeding $40\,\mathrm{dB}$ for TE polarized modes.

	\begin{figure}[htbp]
		\centering
		\centerline{\includegraphics[width=\columnwidth]{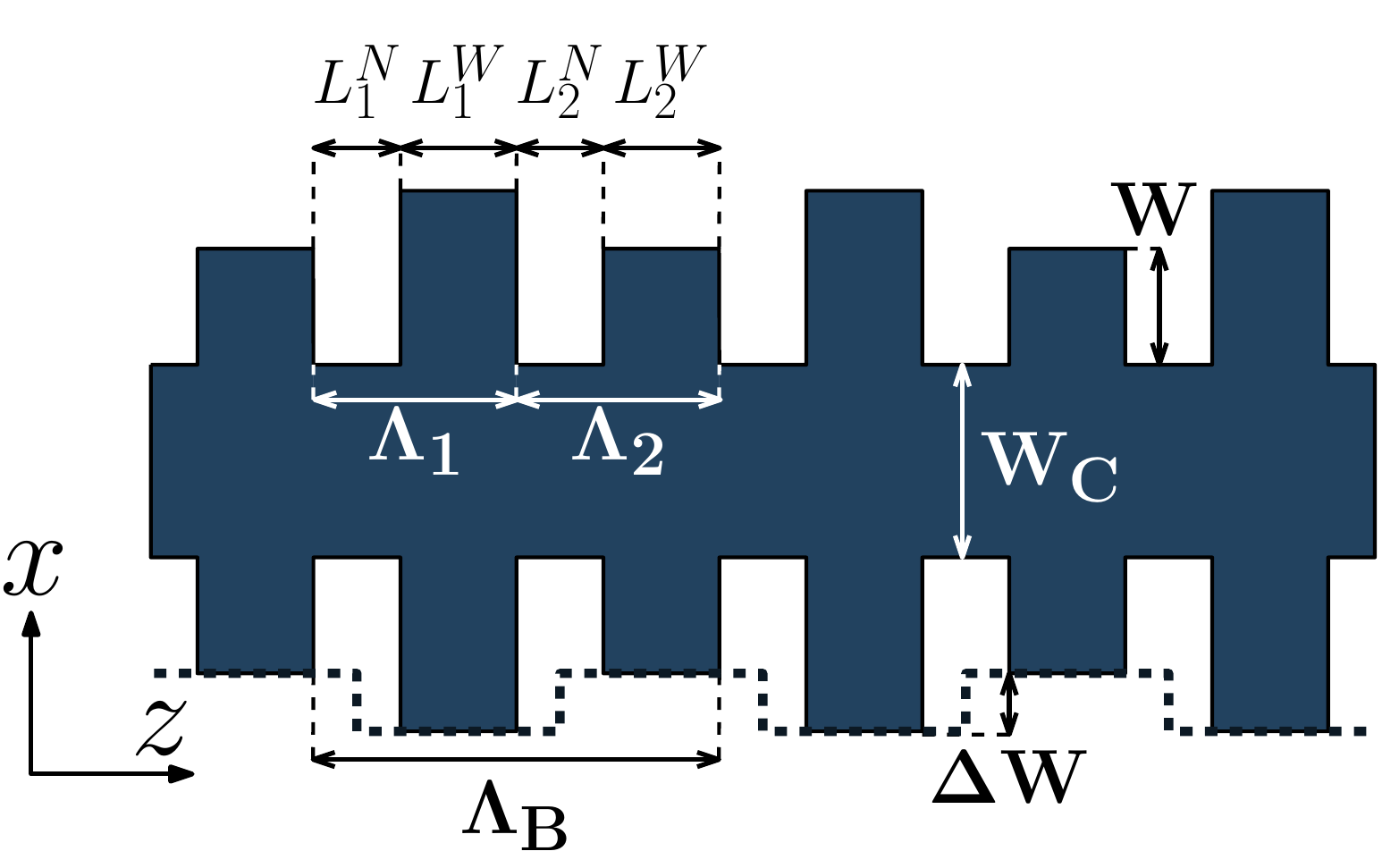}}
		\caption{Schematic of sub-wavelength engineered Bragg filter relying on a double-periodicity, differential width configuration.}
		\label{fig:Schematic}
	\end{figure}

	The bandwidth of the rejection band ($\Delta \lambda$, defined between the first reflection nulls) and the reflectivity ($R$) in a Bragg filter are determined by the coupling coefficient ($k$) between the forward- and backward-propagating modes through the following relations \cite{Formulas}:
	
	\begin{equation}
	\Delta \lambda = \frac{\lambda_{o}^{2}}{\pi n_g} \sqrt{k^2+\left(\frac{\pi}{L_F}\right)^2},
	\label{eq:bandwidth}
	\end{equation}
	
	\begin{equation}
	R = \tanh^2\left(k L_F\right),
	\label{eq:refl}
	\end{equation}
	where $L_F$ represents the filter length, $n_g$ the group index of the (forward- and backward-propagating) waveguide modes, and $\lambda_o$ the central wavelength of the rejection band. Equations (\ref{eq:bandwidth}) and (\ref{eq:refl}) show that narrow-band and high rejection operation can be achieved simultaneously only for long filters with reduced coupling coefficients. While few millimeter long SOI filters (that can be arranged in compact spirals) are easy to implement, small coupling coefficients are difficult to realize. Indeed, due to the high index contrast of Si-wires, small coupling coefficients require very narrow corrugation widths that hinder device fabrication. To overcome this limitation, we propose a sub-wavelength engineered geometry that enables the realization of Bragg filters with small coupling coefficients using substantially larger corrugation widths. As schematically shown in Fig. \ref{fig:Schematic}, our basic filter cell (with a period of $\Lambda_B$) is formed by two rectangularly corrugated sub-cells with sub-wavelength periodicity of $\Lambda_i=L_{i}^{N}+L_{i}^{W}$ (with $i=1,2$).Here, $L_{i}^{N}$ and $L_{i}^{W}$ are the lengths of the narrow and wide sections in each sub-cell. The corrugation widths of the filter are $W_1=W+\Delta W$ and $W_2=W$. Minimum corrugation widths and coupling coefficients, are separately tailored by $W$ and $\Delta W$. Consequently, low coupling coefficients (i.e. very narrow filter width modulations) can be implemented with wide feature corrugations.

	The major advantage of our approach arises from the fact that minimum achievable coupling coefficients are mainly determined by the resolution of the lithography process (that sets minimum size difference between patterns), rather than by the minimum reproducible feature size. For instance, if we consider typical electron-beam (minimum feature size of 50 nm) or deep-ultraviolet (minimum feature size of 100 nm) lithography processes and a resolution of 5 nm, our geometry advantageously provides a ten-fold or even twenty-fold reduction in the minimum implementable modulation width.  
	
	Given the flexibility in the corrugation width design, the minimum feature size of our filter cell is set by the lengths of narrow and wide sections ($L_{i}^{N}$ and $L_{i}^{W}$) required to implement the Bragg periodicity, $\Lambda_B$. The latter is given by 
	
	\begin{equation}
	\Lambda_{\mathrm{B}} = \frac{\lambda_o}{2 n_{BF}},
	\label{eq:Bragg}
	\end{equation}
	where $n_{BF}$ represents the effective index of the Bloch-Floquet mode propagating through the periodic waveguide. Here, we can exploit the sub-wavelength index engineering to reduce waveguide mode index ($n_{BF}$) and enlarge Bragg period, thereby relaxing requirements on the minimum section length. If we set $L_{1}^{N}=L_{1}^{W}=L_{2}^{N}=L_{2}^{W}=L$ and $\lambda_o = 1550\,\mathrm{nm}$, we can implement a Bragg period of $\Lambda_{\mathrm{B}}\sim 340\, \mathrm{nm}$ (with $n_{BF}\sim 2.28$ for $W_C=300\,\mathrm{nm}$ and $W=150\,\mathrm{nm}$), i.e. a minimum section length of $L=85\,\mathrm{nm}$, well within the requirements of our electron-beam lithography. Note that the Bragg period can be further increased to meet minimum feature size requirements of deep-ultraviolet lithography. For instance, a filter with $W_C=170\,\mathrm{nm}$ and $W=150\,\mathrm{nm}$ ($n_{BF}\sim 1.94$) yields a Bragg period of $\Lambda_{\mathrm{B}}\sim 400\, \mathrm{nm}$ and minimum section length of $L=100\,\mathrm{nm}$. 
	
	
	\begin{figure}[htbp]
		\centering
		\centerline{\includegraphics[width=\columnwidth]{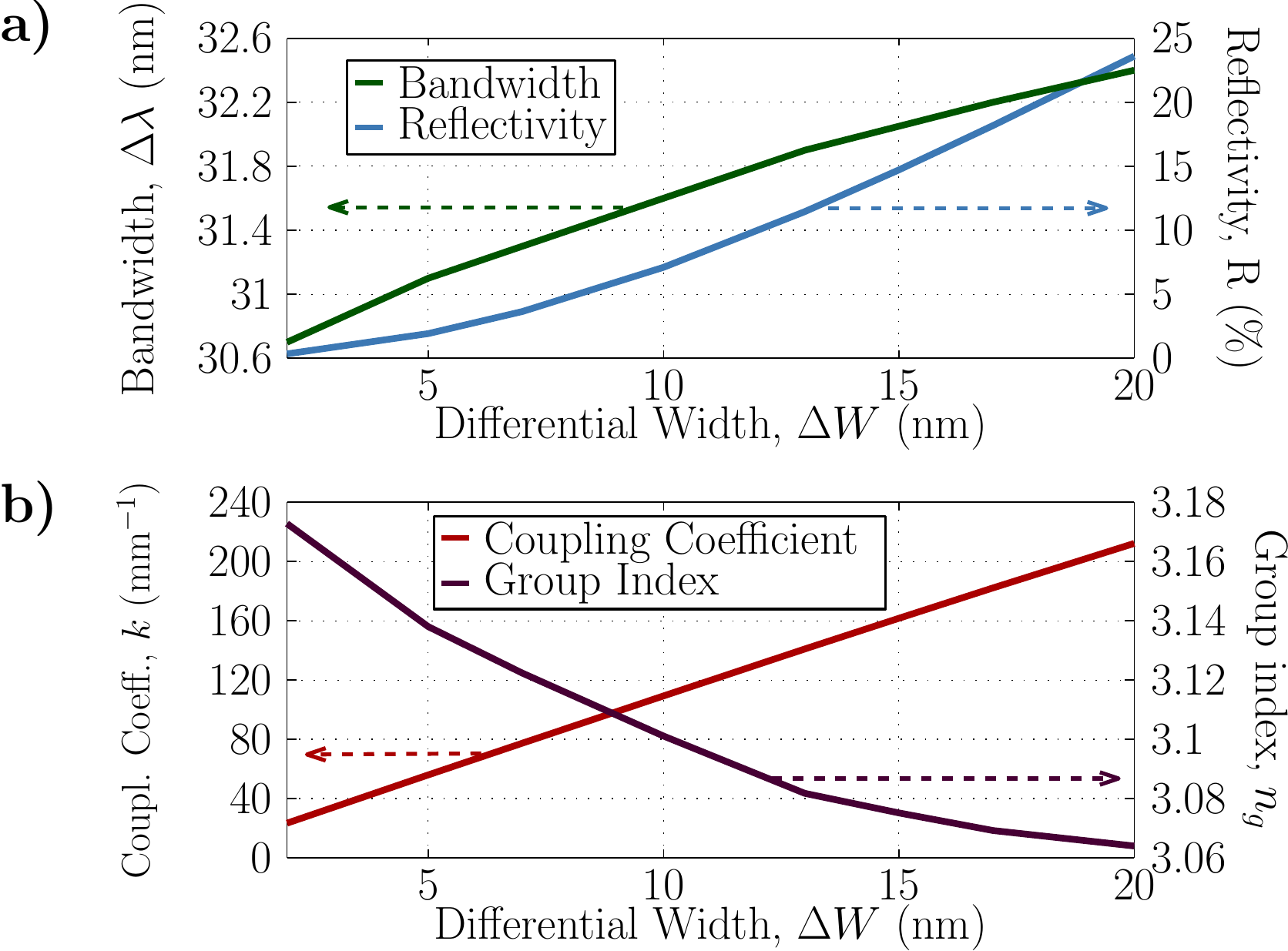}}
		\caption{Calculated (a) reflectivity and bandwidth, and (b) group index and coupling coefficient as a function of the differential width ($\Delta W$) for a double-periodicity Bragg filter with $W_C=300\,\mathrm{nm}$, $W=150\,\mathrm{nm}$ and length of $L_F=25\,\mu\mathrm{m}$.}
		\label{fig:Simulations}
	\end{figure} 
	
	We now analyze the performance of our filter using the 2.5D finite difference time domain (FDTD) simulation tools from Lumerical \cite{Lumerical}. Note that, although less rigorous than a complete 3D simulation, the 2.5D approximation suffices to qualitatively illustrate the operation regime of the proposed Bragg filter with substantially less demanding computation. We have studied the transmission and reflection spectra for a filter length of $L_F=25\,\mu\mathrm{m}$. We include adiabatic transitions between input and output strip waveguides and the filter to minimize off-band loss. We calculate both the reflectivity ($R$) and the rejection bandwidth ($\Delta \lambda$) as a function of the differential filter width ($\Delta W$) for a device having a central strip width of $W_C=300\,\mathrm{nm}$, a minimum corrugation width of $W=150\,\mathrm{nm}$, and lengths of narrow and wide sections of $L_{1}^{N}=L_{1}^{W}=L_{2}^{N}=L_{2}^{W}=85\,\mathrm{nm}$ (filter pitch of $\Lambda_{\mathrm{B}}=340\,\mathrm{nm}$). When $\Delta W=0$, the two filter sub-sections are equal. Hence, the periodic structure has an effective sub-wavelength pitch of 170 nm that suppresses both diffraction and Bragg reflection effects \cite{FirstSwg,SwgReview}. This results in negligible back-reflections $<0.1\%$. Conversely, when a differential width is carried out ($\Delta W \neq 0$), back-reflections arise. As shown in Fig. \ref{fig:Simulations}(a) the bandwidth and the strength of reflectivity are proportional to $\Delta W$. Hence, by designing the physical parameter $\Delta W$, we can tailor the optical properties of the filter. From these values, and using Eq. (\ref{eq:bandwidth}) and (\ref{eq:refl}), we estimate the group index and the coupling coefficient of our filter (see Fig. \ref{fig:Simulations}(b)). These results show that the coupling coefficient varies linearly with the filter differential width.

	\begin{figure}[htbp]
		\centering
		\centerline{\includegraphics[width=\columnwidth]{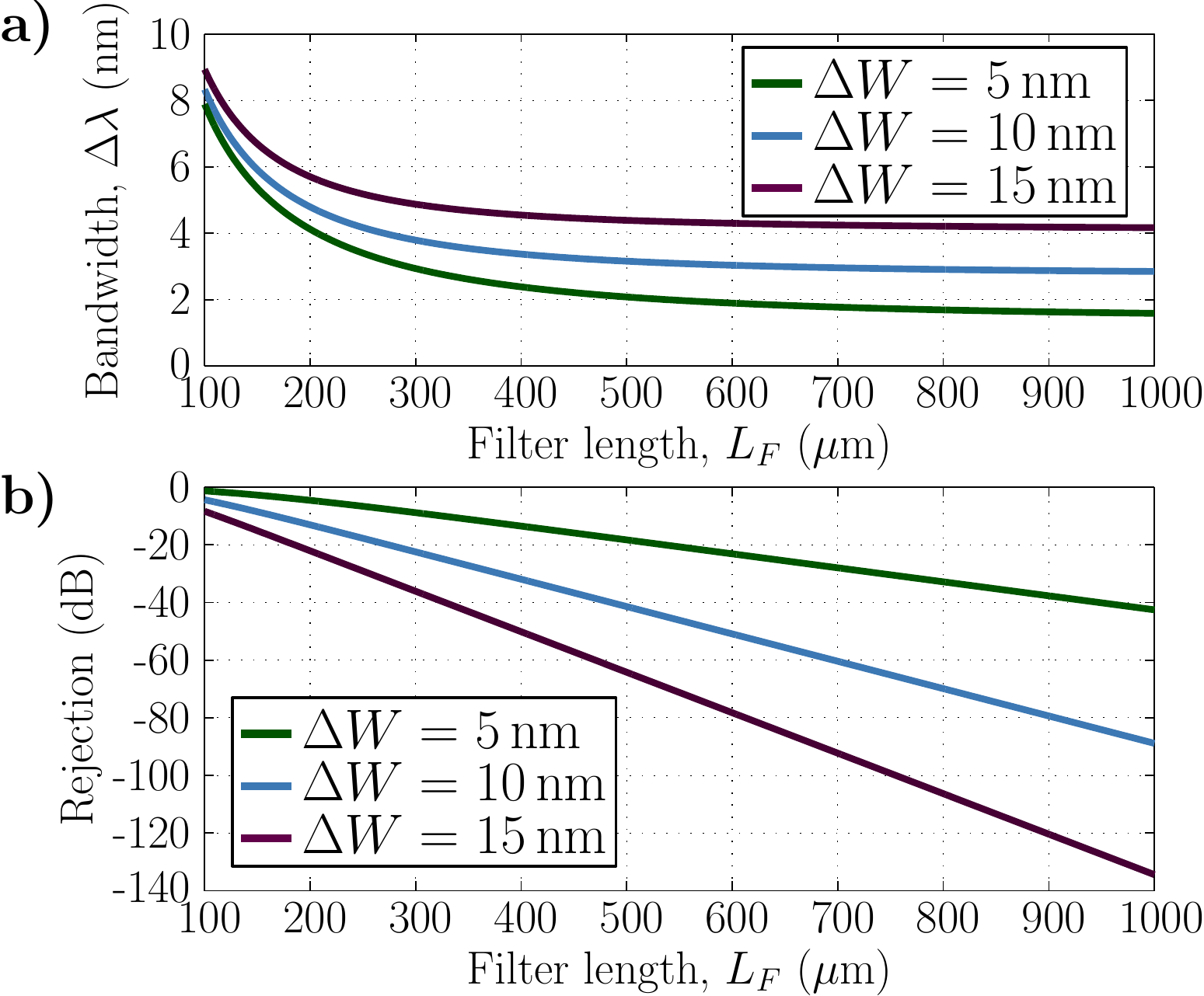}}
		\caption{Estimated (a) rejection bandwidth and (b) transmission level at center wavelength, $\lambda_o$, for double-periodicity Bragg  filter as a function of the length, $L_F$, for various differential widths, $\Delta W$.}
		\label{fig:Calculations}
	\end{figure}  
	
	Using the extracted group index and coupling coefficient and Eq. (\ref{eq:bandwidth}) and (\ref{eq:refl}), we infer the filter rejection bandwidth (Fig. \ref{fig:Calculations}(a)) and the transmission level at the filter central wavelength, $\lambda_o$, (Fig. \ref{fig:Calculations}(b)) as a function of the filter length for differential widths of $\Delta W=5,\,10,\,15\,\mathrm{nm}$. We estimate a bandwidth below $2\,\mathrm{nm}$ and a rejection exceeding $40\,\mathrm{dB}$ for a filter with $\Delta W=5\,\mathrm{nm}$ and $L_F=1000\,\mu\mathrm{m}$.

	To experimentally evaluate the performance of the proposed sub-wavelength engineered geometry, we fabricated the Bragg filters in the SOI platform with a 220 nm thick Si layer and $2\,\mu \mathrm{m}$ thick bottom oxide layer (see Fig. \ref{fig:FabrSEMs}). We set the length of the narrow and wide sections to $L_{1}^{N}=L_{1}^{W}=L_{2}^{N}=L_{2}^{W}=85\,\mathrm{nm}$ (yielding a pitch of $\Lambda_B=340\,\mathrm{nm}$), central strip width to $W_C=300\,\mathrm{nm}$ and minimum corrugation width to $W=150\,\mathrm{nm}$. For comparison, we have designed various differential corrugation widths ($\Delta W=5\,\mathrm{nm},\,10\,\mathrm{nm},\,15\,\mathrm{nm}$) and filter lengths ($L_{F}=100\,\mu\mathrm{m},\,500\,\mu\mathrm{m},\,1000\,\mu\mathrm{m}$). We also have included a $3.4\,\mu\mathrm{m}$ long (ten periods) adiabatic transition, shown in Fig. \ref{fig:FabrSEMs}(a), between the strip waveguide and the filter with minimum corrugation width of $50\,\mathrm{nm}$. The filters have been fabricated using electron beam lithography (Nanobeam NB-4 system, 80 kV) with $5\,\mathrm{nm}$ step-size, followed by a dry etching process with an inductively coupled plasma etcher (SF$_6$ gas) to pattern the structures. Devices are covered with PMMA for protection.

\begin{figure}[htbp]
		\centering
		\centerline{\includegraphics[width=\columnwidth]{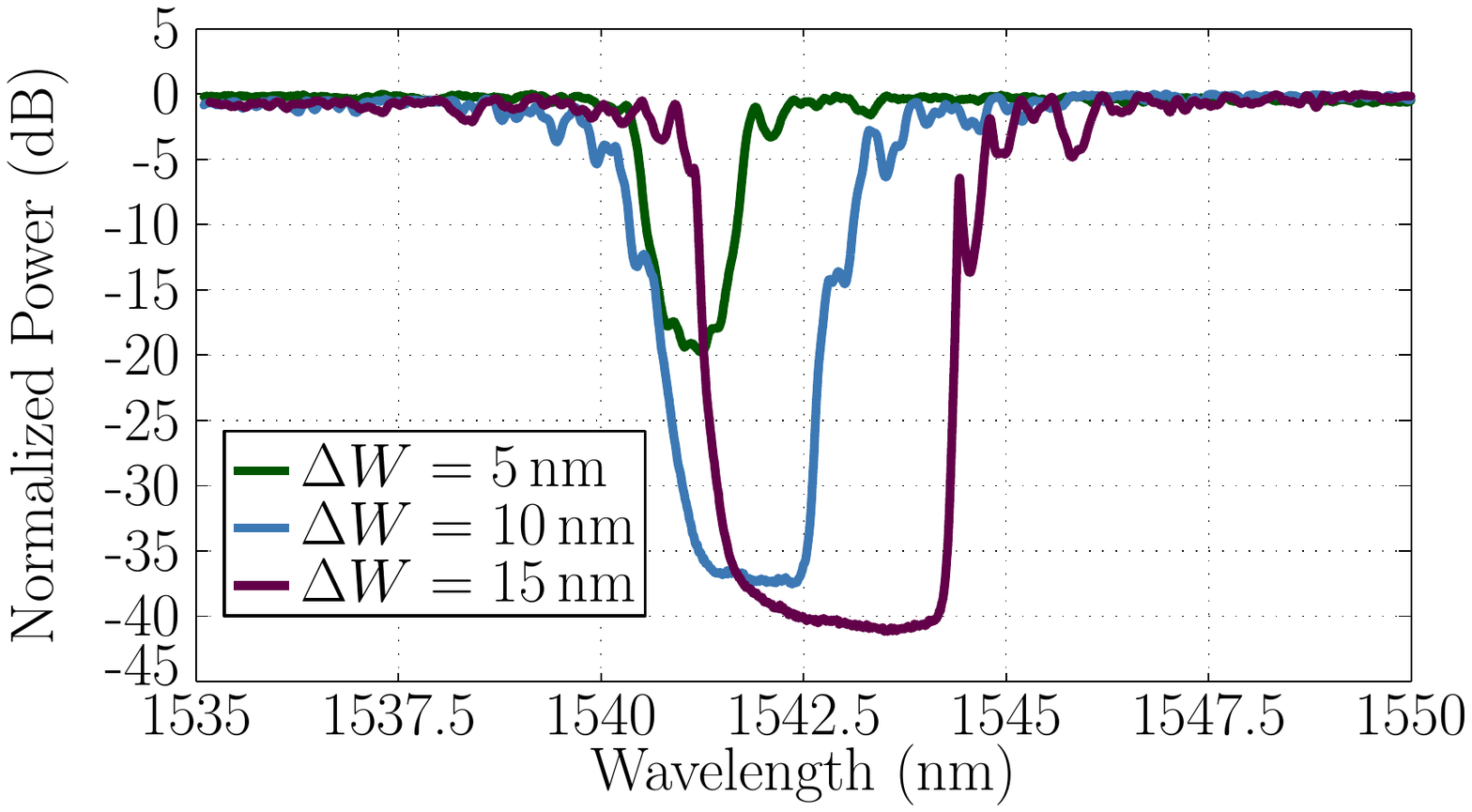}}
		\caption{Measured transmission spectrum of double-periodicity Bragg filters with length of $L_F=500\,\mu\mathrm{m}$ for various differential widths ($\Delta W$).}
		\label{fig:DeltaWEffect}
	\end{figure}

	Light is injected and extracted through the chip surface using grating couplers and cleaved single mode (SMF-28) optical fibers. Sub-wavelength engineered grating couplers \cite{SWG_Coupler_A,SWG_Coupler_B} were optimized to couple TE-polarized light with reduced Fabry-P\'{e}rot ripples that could distort the filter response \cite{RippleEffect}. Figure \ref{fig:DeltaWEffect} shows the measured transmission spectra as a function of the Bragg corrugation width for a filter length of $L_F=500\,\mu\mathrm{m}$. As expected, wider Bragg corrugations result in wider and deeper rejection bands. Note that wider corrugation widths also yield higher (Bloch-Floquet) mode effective indices that redshift the central wavelength of the rejection band. The differences between calculated and experimental rejection values for wider filter corrugations arise from the fabrication errors that limit the maximum achievable rejection level to about 40 dB.

\begin{figure}[htbp]
		\centering
		\centerline{\includegraphics[width=\columnwidth]{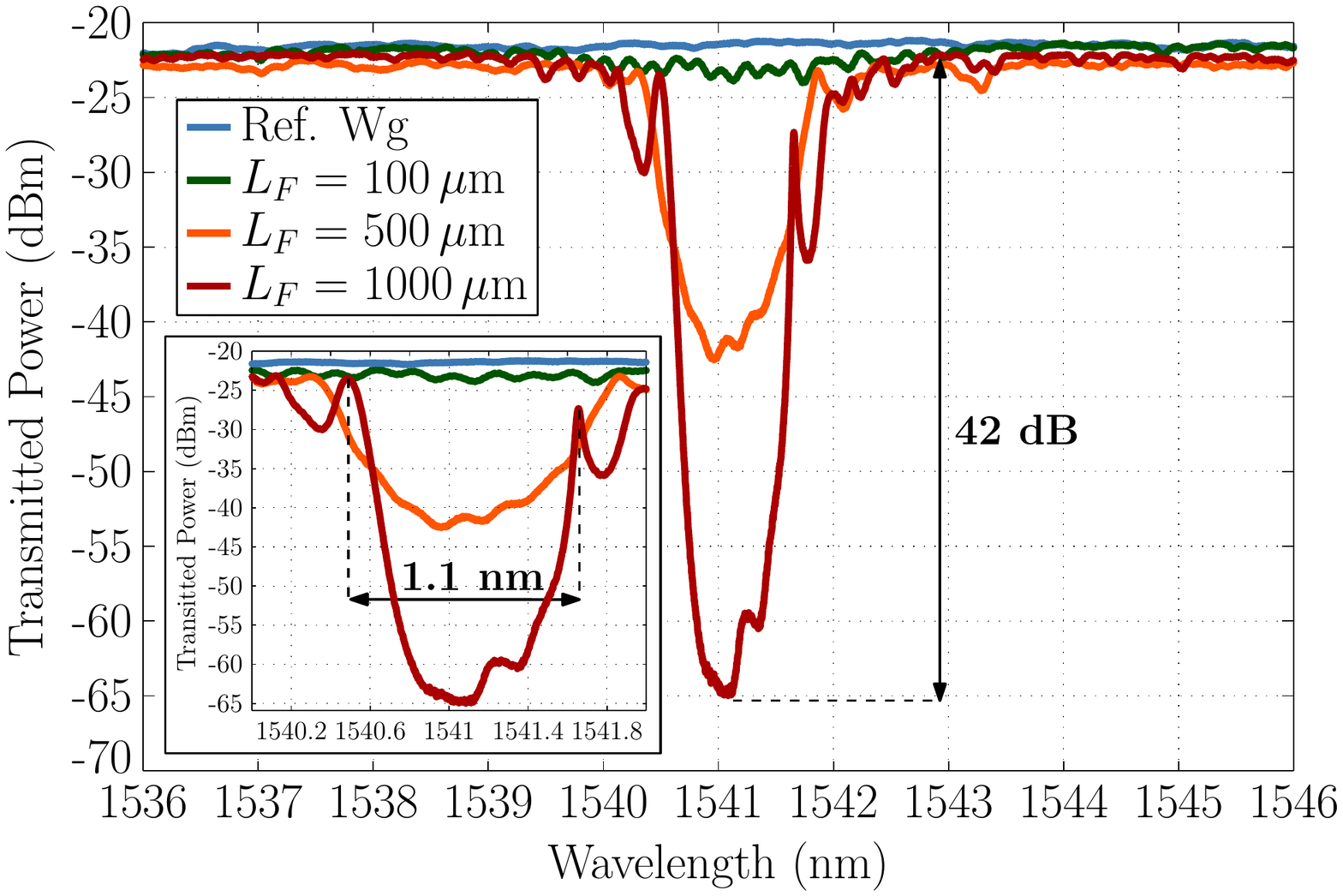}}
		\caption{Measured transmission spectrum of reference strip waveguide of $1000\,\mu\mathrm{m}$ length and double-periodicity filters with differential width of $\Delta W=5\,\mathrm{nm}$ and different filter lengths, $L_F$.}
		\label{fig:LWEffect}
	\end{figure}

	Figure \ref{fig:LWEffect} shows the measured transmission spectra as a function of the filter length ($L_F$) for a differential width of $\Delta W=5\,\mathrm{nm}$. The filter with a length of $L_F=1000\,\mu\mathrm{m}$ exhibits a remarkably narrow bandwidth of $1.1\,\mathrm{nm}$, with a rejection level exceeding $40\,\mathrm{dB}$. This is almost a $30\,\mathrm{dB}$ rejection improvement compared to previously reported sub-wavelength engineered Bragg filters with comparable lengths \cite{SwgBraggExp}. The minimum corrugation width of our filter cell, of $W=150\,\mathrm{nm}$, is ten times wider than conventional single-etch Bragg filters \cite{Small_Teeth} and two times wider than TM-polarized Bragg filters \cite{Spiral_TM} with similar bandwidths. Figure \ref{fig:LWEffect} also shows the transmission spectrum of a reference waveguide featuring a $450\,\mathrm{nm}$ width (average between the narrow and wide filter sections) and a length of $1000\,\mu\mathrm{m}$. The difference in the off-band transmission levels of the different Bragg filters and the reference waveguide is within the alignment precision of our setup. This therefore shows that our filter exhibits low loss for wavelengths outside the stop band.  
	
	In summary, we have reported for the first time, the design and experimental demonstration of a sub-wavelength engineered Bragg filters relying on a differential width geometry. We relax the minimum width constraints by leveraging the differential width between two filter sub-periods, while using sub-wavelength index engineering to overcome longitudinal minimum feature limitations. Exploiting this concept, we have experimentally demonstrated a remarkably narrow bandwidth of $1.1\,\mathrm{nm}$ with rejection exceeding $40\,\mathrm{dB}$ for a filter cell with minimum transversal and longitudinal features of $150\,\mathrm{nm}$ and $85\,\mathrm{nm}$, respectively. These results are an important step towards the realization of new generation, high performance, quantum chips, integrating on a single SOI substrate photon-pair sources and pump-rejection filters.
	
	\section*{Funding Information}
	This work was partially funded by the Agence Nationale de la Recherche (ANR-SITQOM-15-CE24-0005), the European Research Council (ERC) under the European Union's Horizon 2020 research and innovation program (ERC POPSTAR grant agreement No 647342) and the NANO2017 programme with STMicroelectronics funded by the “Minist\`ere de l'\'economie, de l'industrie et du num\'erique”, “D\'el\'egation G\'en\'erales des Entreprises”.

\end{document}